\documentclass[a4paper, 10pt,twocolumn]{article}

 \usepackage[dvips]{graphicx}
 \usepackage{amssymb} 
 \usepackage{amsmath} 
\begin{document}

%% ------------------------------------------------------------------------ %%
%
% TITLE
%
%% ------------------------------------------------------------------------ %%

\title{Vlasov-Poisson simulations of electrostatic parametric instability for localized Langmuir wave packets in the solar wind}

\author{P.~Henri$^{1,2}$, F.~Califano$^{2,1}$, C.~Briand$^{1}$, A.~Mangeney$^{1}$ \\
\small{$^{1}$ LESIA, Observatoire de Paris, CNRS, UPMC, Universit\'e Paris Diderot; 5 Place J.~Janssen, 92190 Meudon, France} \\
\small{$^{2}$ Dip. Fisica, Universit\`{a} di Pisa; Largo Pontecorvo 3, 56127 Pisa, Italy}  \\
 }

\date{}

%% ------------------------------------------------------------------------ %%
% ABSTRACT
%
%% ------------------------------------------------------------------------ %%

\twocolumn[
  \begin{@twocolumnfalse}
    \maketitle
\begin{abstract}
% Context
Recent observation of large amplitude Langmuir waveforms during a Type~III event in the solar wind have been interpreted as the signature of the electrostatic decay of beam-driven Langmuir waves. This mechanism is thought to be a first step to explain the generation of Type~III radio emission. 
% Aim and Methods
The threshold for this parametric instability in typical solar wind condition has been investigated through 1D-1V Vlasov-Poisson simulations.
% Results
We show that the amplitude of the observed Langmuir beat-like waveforms is of the order of the effective threshold computed from the simulations. The expected level of associated ion acoustic density fluctuations have also been computed for comparison with observations. \\
%}
    \end{abstract}
  \end{@twocolumnfalse}
]

%% ------------------------------------------------------------------------ %%
% TEXT
%
%% ------------------------------------------------------------------------ %%

%%%%%%%%%%%
\section{Introduction}
%%%%%%%%%%%

% Position du probleme: Description of solar wind observations :
%-------------------------------------------------------------------------------------------

During a solar flare, high energy electrons (1-100 keV) are expelled from the solar corona and travel
along the interplanetary magnetic field lines, producing a bump on the local electron distribution function. Langmuir waves then grow via the so-called Óbump-on-tail instability. Langmuir waves are observed at amplitude large enough ($E^2/(8 \pi n T) \simeq 10^{-3} - 10^{-4}$) to further excite electromagnetic waves through non-linear processes. These electromagnetic waves are the main signature of Type~III radio bursts.
Wave-wave interaction through parametric instabilities have been shown to be the physical mechanism underlying the non linear evolution of large amplitude Langmuir waves. 

Langmuir electrostatic decay (hereafter LED) -- also called Langmuir decay instability (LDI) or parametric decay instability (PDI) in the literature -- enables energy transfer from a finite amplitude Langmuir wave $L$ toward a second Langmuir wave $L$' and an ion acoustic wave (hereafter IAW) $S$ through a three-wave resonant interaction:
\[  L \rightarrow L' + S  \]
This process is thought to be a first step toward the generation of solar wind type~III radio emission at twice the plasma frequency ($T_{2 f_{pe}}$), as a result of a coalescence of the two Langmuir waves  \cite{ginzburg58}: 
\[  L~+~L'~\rightarrow~T_{2 f_{pe}}   \]

Spectral observations of high frequency intense Langmuir waves and low-frequency ion acoustic waves during Type~III bursts have been interpreted as a signature of Langmuir electrostatic decay \cite{Lin&al1986ApJ}. Furthermore, waveform observations during Type~III bursts of modulated Langmuir wave packets on one hand \cite{Cairns&Robinson1992GRL, hospodarsky&gurnett1995GRL, Bale&al1996GRL, Li&al2003PhPl} and of IAW frequency drift associated with Type~III electron beam velocity drift on the other hand \cite{Cairns&Robinson1995ApJ}, have both been interpreted by the Langmuir electrostatic decay.

% Description of Electrostatic langmuir decay + bibliography:
%------------------------------------------------------------------------------------
The LED is a resonant parametric instability. 
To be resonant, the three-wave coupling requires the fundamental equations of energy and momentum conservation to be satisfied:
\begin{equation} \label{eq:resonantconditions}
	\omega_L = \omega_{_{L'}}+ \omega_S  \; \; \; \; \; \; \; \; \; \;  
	\vec{k}_{_{L}}  = \vec{k}_{_{L'}} + \vec{k}_S  
\end{equation} 
where $\omega$ and $\vec{k}$ are the frequency and wavenumber of the waves.
Moreover, for the LED to develop, the electric field of the mother Langmuir wave has to reach a critical value.
The analytical calculation of this threshold has been discussed in \cite{NishikawaJPSJ1968,Sagdeev&Galeev1969,Dysthe&Franklin1970PlasmaPhysics,Bardwell&Goldman1976Apj,Robinson&al1993ApJ} by considering three monochromatic waves. The underlying idea is that for the instability to develop, the growth rate $\gamma_{_{LED}}$ of the two product waves must be higher than their own linear Landau damping rates $\gamma_{_{L'}}$ and $\gamma_{_{S}}$:
\[
	\gamma_{_{LED}} > \sqrt{\gamma_{_{L'}} \gamma_{_{S}}}
\]
The threshold for parametric decay of the mother Langmuir wave is thus evaluated in term of the electric to kinetic energy ratio \cite{Bardwell&Goldman1976Apj}:
\[
	\frac{\epsilon_{0} E^{2}}{n k_{B} T} > 8 \frac{\gamma_{_{L'}}}{\omega_{_{L'}}} \frac{\gamma_{_{S}}}{\omega_{_{S}}} 
\]
with $\omega_{L'}$, $\omega_S$ the angular frequency of the daughter waves. The threshold for electrostatic decay has been estimated for typical solar wind parameters to be $(\epsilon_{0} E^{2})/(n k_{B} T) \geq 2.5 \times 10^{-5}$ \cite{lin86a}. 
To our knowledge, only LED that couples three monochromatic waves have been considered analytically. Indeed, the analytical treatment of resonance becomes complicated when considering a finite frequency bandwidth, in which case a numerical treatment is necessary.

Earlier related works on simulations of LED have been based on the Zakharov equations \cite{Sprague&Fejer1995JGR,Gibson&al1995PhRvE,Soucek&al2003AnGeo}, considering the instability as a fluid mechanism. Recently, kinetic simulations using PIC codes \cite{Matsukiyo&al2004JGR,Huang&Huang2008ChA&A} and Vlasov codes \cite{goldman96, Umeda&Ito2008PhPl} have shown that the beam-plasma interaction can saturate through LED and have been able to reproduce the modulated shape of Langmuir waves. 
 
If the threshold for LED is commonly thought to be at moderate amplitude, a recent numerical work \cite{Umeda&Ito2008PhPl} claims that no LED would occur until the electric energy is comparable to the plasma kinetic energy. As a result, it would be no more possible to consider the LED as a mechanism for the generation of Type~III radio emission.

% MON probleme:
%-------------------------
\cite{Henri&al2009JGR}, hereafter Paper I, recently reported direct observations of Langmuir waves decaying into secondary Langmuir waves and acoustic waves during a Type~III solar event, from STEREO/WAVES data. They found that the Doppler-shifted frequencies of the three observed waves satisfy the resonant relations of momentum and energy conservation expected for three-wave coupling. A bicoherence analysis confirmed the phase coherence of the three waves. 
In this former work, the LED threshold and the growth rate of IAW density fluctuations generated by LED were both evaluated from analytical solutions involving a purely monochromatic three-wave coupling \cite{Sagdeev&Galeev1969}.
However, observations show that: (i) the large amplitude Langmuir waves are isolated wave packets with a packet width of the order of a few wavelengths; (ii) ion and electron temperatures are close ($ 0.1 < T_{e} / T_{i} < 10$) so that ion acoustic waves associated with the LED should be Landau-damped.
Thus several questions remain open about the generation and the dynamics of the LED observed in the solar wind:
\begin{enumerate}
	\item What are the threshold and the growth rate of the LED when electron and ion temperatures are close? What is the effective threshold of LED when the mother Langmuir wave is a localized wave packet?
	\item What is the saturation level for IAW density fluctuations in these conditions? What is the expected level of IAW density fluctuations where saturation is not observed?
\end{enumerate}

The goal of this paper is to answer these questions by studying the dynamics of the LED through 1D-1V Vlasov-Poisson simulations. \\

% Plan du travail:
%----------------------
The paper is organized as follow. The Vlasov-Poisson simulation model is described in section~\ref{section:model}. The numerical results are presented in section~\ref{section:simulationresults}, first for a monochromatic mother Langmuir wave (sect.~\ref{section:monochromatic}), then for a mother Langmuir wave packet (sect.~\ref{section:wavepacket}). Growth rates, saturation levels for IAW density fluctuations and threshold for LED are studied. The simulation results are finally discussed in section~\ref{section:discussion} in the context of solar wind observations. Particular attention is paid to the case of equal temperature for electrons and ions, in which case the growth of IAW should be limited by its Landau damping.

%%%%%%%%%%%%%%%%%%%%%%%%
\section{Full Vlasov-Poisson simulation model} \label{section:model}
%%%%%%%%%%%%%%%%%%%%%%%%
 
 % justification methode Vlasov-Poisson:
%------------------------------------------------------
In typical solar wind conditions, the ratio between electron and ion temperatures fluctuates around 1. The IAW produced during three-wave coupling should then be suppressed by Landau damping. This would limit the development of the IAW and thus the LED. However, the IAW Landau damping rate in such temperature condition cannot be solved analytically, so that it cannot be included in a fluid code. Kinetic effects due to wave-particle interactions are to be taken into account self-consistenly as a possible limitation for the instability to grow. 
A Vlasov-Poisson approach has thus been used to study LED dynamics. It allows to consider self-consistently the decay of the Langmuir wave together with damping effect on the product waves.
Since solar wind electrons are weakly magnetized ($\omega_{ce} / \omega_{pe} \simeq 10^{-2}$), magnetic effects are discarded. \\

% Description du modele:
%----------------------------------
The Vlasov-Poisson system is solved for the electron and ion distribution function, $f_{e}(x,v,t)$ and $f_{i}(x,u,t)$, with the numerical scheme described in \cite{Mangeney&al2002JCoPh}, limiting our study to the 1D-1V case.
The equations are normalized by using the following characteristic electron quantities: the
charge $e$, the electron mass $m_{e}$, the electron density $n_{e}$, the plasma (angular) frequency $\omega_{pe} = \sqrt{4 \pi n_{e} e^2 / m_{e}}$, the Debye length $\lambda_{D} = \sqrt{T_{e} / 4 \pi n_{e} e^2}$, the electron thermal velocity $v_{th,e} = \lambda_{D} \omega_{pe} = \sqrt{T_{e} / m_{e}}$ and an electric field $\bar E = m_{e} v_{th,e} \omega_{pe} / e$. Then, the dimensionless equations for each species read:

\begin{equation} \label{eq:vlasovelectrons}
\frac{\partial f_{e}}{\partial t} + v \frac{\partial f_{e}}{\partial x} - (E+E_{ext}^{init}) \frac{\partial f_{e}}{\partial v} = 0 
\end{equation}

\begin{equation} \label{eq:vlasovions}
\frac{\partial f_{i}}{\partial t} + u \frac{\partial f_{i}}{\partial x} + \frac{1}{\mu} (E+E_{ext}^{noise}) \frac{\partial f_{i}}{\partial u} = 0
\end{equation}

\begin{equation} \label{eq:poisson}
\frac{\partial^2 \phi}{\partial x^2} = \int f_{e} dv - \int f_{i} du \\ \mathrm{\ ; \ } \\ E = - \frac{\partial \phi}{\partial x}
\end{equation}

\noindent where $v$ (resp. $u$) is the electron (resp. ion) velocity normalized to the electron thermal velocity. Here $\mu=m_{i} / m_{e} =1836$ is the ion-to-electron mass ratio. $\phi$ and $E$ are the self-consistent electric potential and electric field generated by the plasma charge density fluctuations according to Poisson equation (Eq.~\ref{eq:poisson}). $E_{ext}^{init}$ and $E_{ext}^{noise}$ are 'external' drivers added to the Vlasov equation that can be switched on or off during the run. 
The electron (resp. ion) distribution function is discretized in space for $0 \leq x < L_{x}$, with $L_{x} = 5000\ \lambda_{D}$ the total box length, with a resolution of $dx = \lambda_{D}$.
The electron velocity grid ranges over $-5\ v_{th,e} \leq v \leq +5\ v_{th,e}$, with a resolution of $d v = 0.04\ v_{th,e}$. 
(resp. $-5\ u_{th,i} \leq u \leq +5\ u_{th,i}$, with a resolution of $d u = 0.04\ u_{th,i}$ for the ion velocity grid, where $u_{th,i}$ is the ion thermal velocity). 
Finally, periodic boundary conditions are used in the spatial direction.

In all the runs the following initial conditions have been taken: electron and ion distributions functions are initially Maxwellian with respect to velocity, with a random noise in density:
\begin{equation} \label{eq:condinitelectrons}
f_{e}(x,v) = \frac{e^{-v^2} }{\sqrt{2 \pi}}  [1 + \epsilon \sum_{k} cos(kx+\psi_{k}) ]
\end{equation} 
\begin{equation} \label{eq:condinitions}
f_{i}(x,u) = \frac{\sqrt{\frac{\Theta}{\mu}} e^{-\frac{\Theta}{\mu} u^2} }{\sqrt{2 \pi}}  [1 + \epsilon' \sum_{k} cos(kx+\psi'_{k}) ]
\end{equation}
where $\Theta = T_{i} / T_{e}$ is the ion-to-electron temperature ratio set to $\Theta = 0.1$ or 1 in the different runs. $\psi_{k}$ and $\psi_{k}'$ are random phases with a uniform distribution. The parameter $\epsilon$ (resp. $\epsilon'$) is the amplitude of the initial electron (resp. ion) density level of noise. 
The parameters ($\epsilon$,$\epsilon^{\prime}$) are chosen so that the development of the instability happens relatively rapidly, as Vlasov codes have a very low level of numerical noise. 
The influence of the level of initial density fluctuation on the LED has been studied for values $10^{-8} < \epsilon, \epsilon' < 10^{-4}$. Neither the growth rate of density fluctuations, nor its saturation level are modified by the choice of the parameters $\epsilon$ and $\epsilon'$, as long as they remain weak. Only the time needed for the instability to saturate is modified. The chosen values $\epsilon = \epsilon' = 10^{-5}$ are a  good compromise to limit the computation time. 
However, when the electron and proton temperatures are of the same order the initial perturbation on ion density is rapidly damped out and the generation of the IAW starts from the numerical noise. When starting from a monochromatic Langmuir wave (section~\ref{section:monochromatic}), the interaction time between the waves is 'infinite'. In this case the IAW can grow from the numerical noise to significant values providing we wait for long enough. On the other hand, when starting from a Langmuir wave packet (section~\ref{section:wavepacket}) the interaction time between the waves is now finite. A continuous injection of noise is then needed to seed the instability. 
This is why an external driver~$E_{ext}^{noise}$ is added in the Vlasov equation for ions (Eq.~\ref{eq:vlasovions}). The aim is to control the generation of a continuous incoherent noise in the proton density. For self-consistency, the driver is used in both cases $\Theta = 0.1$ and $\Theta = 1$. The details of the forcing are described in appendix~\ref{appendix:Eext_ions}. From a physical point of view, IAW grow from the proton density fluctuations background. Note that such density irregularities are observed in the solar wind \cite{Celnikier&al1987A&A,Harvey&al1988AdSpR}. \\

In space conditions, the full physical process for the nonlinear evolution of large amplitude Langmuir waves is characterized by two successive steps. The first one is the generation of Langmuir waves from the bump-on-tail instability, the second one the electrostatic decay of these Langmuir waves if their amplitudes reach the LED threshold. 
Other authors have already studied the generation of beam driven Langmuir wave packets \cite{Omura&al1994GRL,Omura&al1996JGR,Silin&al2007PhPl_b,UmedaNPG2007}. Among their results, they have shown that the beam-driven Langmuir waves are localized packets. This localization is explained through the mechanism of kinetic localization in the framework of a nonlinear trapping theory \cite{Muschietti&al1995JGR,Muschietti&al1996JGR,Akimoto&al1996PhPl,Usui&al2005JGR}. 
It is has also been shown that the beam-driven Langmuir wave packets are formed on time scales considerably shorter than those of parametric instabilities \cite{Intrator&al1984PRL,Akimoto&al1996PhPl}. 

Our aim here is to focus the attention on the LED process. 
Motivated by the fact that the generation and localization of beam-driven Langmuir waves are decoupled from the LED process, we choose not to generate Langmuir waves by a bump-on-tail instability, but to resonantly grow the initial Langmuir wave by means of an electric field $E_{ext}^{init}$ added in Eq.~\ref{eq:vlasovelectrons}. 
This method enables to have a direct control on the energy and the spectrum of the initial Langmuir wave and avoid other effects due to the non-Maxwellian character of the initial distribution function. Switched on at the beginning of the run, it acts as a driver to resonantly grow the wave with the desired spectrum and electric field amplitude. Details on the external driver~$E_{ext}^{init}$ are given in Appendix~\ref{appendix:Eext}. The external driver is then switched off and the generated Langmuir wave evolves self-consistenly. In section~\ref{section:monochromatic}, a monochromatic Langmuir wave of wavelength~$\lambda_{_{L}}$ and amplitude~$E_{_{L}}$ is generated by imposing $E_{ext}^{init} = E_{ext}^{(1)}$ (cf. Eq.~\ref{eq:Eext_monochromatic}). In section~\ref{section:wavepacket}, a Langmuir wave packet of mean wavelength~$\lambda_{_{L}}$, packet width~$\Delta$ and maximum amplitude~$E_{_{L}}$ is generated by imposing $E_{ext}^{init} = E_{ext}^{(2)}$ (cf. Eq.~\ref{eq:Eext_wavepacket}). The external 
electric field $E_{ext}^{init}$ is switched off when the amplitude (resp. maximum amplitude) of the generated Langmuir wave (resp. wave packet) reaches the desired value $E_{_{L}}$. This happens typically for time $t < 300\ \omega_{pe}^{-1}$, small compared to the decay time scale. So, the resonant generation of the Langmuir wave (resp. wave packet) does not interfere with the LED mechanism.

% Observational parameters:
In order to compare the simulation results with observations of electric waveforms during Type~III events, the Langmuir wavelength and amplitude are set as indicated by solar wind observations.
Langmuir waves grow by resonance with an electron beam at phase velocity $v_{_{L}}^{\Phi} = \omega_{_{L}} / k_{_{L}} \simeq V_{beam}$.  Typical electron beams span in the range $0.05 - 0.2\ c$ \cite{Dulk&al1987A&A,Hoang&al1994A&A}. Taking into account the Langmuir dispersion relation, we deduce Langmuir wavelengths in the range $\lambda_{_{L}} = [100 - 600]\ \lambda_{D}$.
However, for a temperature ratio $\Theta = 1$ (resp. $\Theta = 0.1$), the daughter Langmuir wave packet is expected, from Eq.~\ref{eq:resonantconditions} and respective dispersion relations, to be backscattered for $\lambda_{_{L}} < 285\ \lambda_{D}$ (resp. $\lambda_{_{L}} < 385\ \lambda_{D}$), and scattered forward for $\lambda_{_{L}} > 285\ \lambda_{D}$ (resp. $\lambda_{_{L}} > 385\ \lambda_{D}$). Thereby, mother and daughter Langmuir Doppler-shifted frequencies are very close and hardly separated on observation. 
Thus, to be able to compare the simulations with observations, we choose Langmuir wavelengths that allow observations of LED, namely $\lambda_{_{L}} = 100\ \lambda_{D}$ for the initial monochromatic Langmuir wave in section~\ref{section:monochromatic}, and $ 50\ \lambda_{D} < \lambda_{_{L}} < 400\ \lambda_{D}$ for the initial Langmuir wave packet in section~\ref{section:wavepacket}.
Since the phase velocities of the two expected Langmuir waves are large enough, their Landau damping can be neglected. 
Thus the phase velocity of the Langmuir waves do not need to be resolved in the electron distribution function velocity box.
On the other hand, the expected IAW travel at the ion-sound speed $c_{s} = \sqrt{(T_{e}+T_{i})/m_{i}}$, which is of the order of the ion thermal speed when electron and ion temperature are close, thus leading to a large Landau damping. The ion-sound speed is resolved in the ion distribution function velocity box.

The typical Langmuir waves amplitude is directly given by waveform observations and normalized to the electron temperature:
\begin{equation} \label{eq:normalizedelectricfield}
E = \sqrt{\frac{\epsilon_{0} E_{obs}^2 /2}{n k_{b} T_{e}}}
\end{equation} 
\noindent In Paper I, the typical observed values are $10^{-2} < E_{_{L}} < 10^{-1}$. We choose several amplitudes in the range $10^{-3} < E_{_{L}} < 1$.

%########################
\section{Numerical results} \label{section:simulationresults}

During the first part of the simulation ($0 < t < t_{0}$) the external electric pump generates a Langmuir monochromatic wave (resp. wave packet) at the desired amplitude. At time $t = t_{0}$, the external pump is switch off when the Langmuir wave (resp. wave packet) reaches the desired value. For $t > t_{0}$ the Langmuir wave evolves self-consistently. We consider time $t=t_{0}$ as the beginning of the numerical experiment.

The typical intensity of the electric field of the IAW generated by LED is much below the level of Langmuir electric noise. In the following, the ion \emph{density} is therefore used as a tracer for IAW. 

%%%%%%%%%%
 \begin{table*}
 \begin{center}
 \begin{tabular}{|c|c|c|c|c|c|}
\hline
                        &   Waves  &  Wavelength     &  Wavenumber &  Phase velocity &  Group velocity  \\
                        &          & ($\lambda_{D}$) & ($\lambda_{D}^{-1}$) & ($v_{th}^{e} = \lambda_{D} \omega_{pe}^{-1}$) & ($v_{th}^{e} = \lambda_{D} \omega_{pe}^{-1}$) \\
\hline
$\Theta = 0.1$ & $L$   & $\lambda_{_{L}} =  100$ & $k_{_{L}} = 0.063$ & $v^{\phi}_{_{L}} = 16$ & $v^{g}_{_{L}} = 0.189$  \\
                        & $L'$  & $\lambda_{_{L'}} = 132 $ & $k_{_{L'}} = (-) 0.047$ & $v^{\phi}_{_{L'}} = - 21$ & $v^{g}_{_{L'}} = -0.141$ \\
                        & $S $ & $\lambda_{_{S}} =  57  $ & $k_{_{S}} = 0.110$ & $c_{s} =  0.024$ & $c_{s} =  0.024$ \\
\hline
$\Theta = 1$    & $L$   & $\lambda_{_{L}} = 100$  &  $k_{_{L}} = 0.063$ & $v^{\phi}_{_{L}} = 16$ & $v^{g}_{_{L}} = 0.189$  \\
                        & $L'$  & $\lambda_{_{L'}} = 154$ & $k_{_{L'}} = (-) 0.041$ & $v^{\phi}_{_{L'}} = - 24.5$ & $ v^{g}_{_{L'}} = -0.123$ \\
                        & $S $  & $\lambda_{_{S}} = 60 $ &  $k_{_{S}} = 0.104$ & $c_{s} = 0.033$ & $c_{s} =  0.033$ \\
\hline
\end{tabular}
 \caption{Expected waves involved in the Langmuir electrostatic decay, given the initial Langmuir wavelength and using equations~\ref{eq:resonantconditions} for $\Theta = 0.1$ (top line) and $\Theta = 1$ (bottom line). It shows for each wave respectively the expected wavelength (in Debye length $\lambda_{D}$), corresponding wavenumber (in inverse Debye length $\lambda_{D}^{-1}$) used in the paper figures, phase and group velocities (in $\lambda_{D} \omega_{pe}^{-1}$)} \label{table:expectedwaves}  
\end{center}
\end{table*} 
%%%%%%%%%%

%-------------------------------------------
\subsection{Electrostatic decay of a monochromatic Langmuir wave} \label{section:monochromatic}

In this section, we consider the evolution of a monochromatic wave generated by the external pump defined in Eq.~\ref{eq:Eext_monochromatic}. Table~\ref{table:expectedwaves} summarizes the expected wavelength, wavenumbers, phase velocity and group velocity in the case of three monochromatic resonant waves, as a function of the mother Langmuir wavelength $\lambda_{_{L}}$ and the temperature ratio $\Theta$. \\

\begin{figure*}
\begin{center}
\noindent \includegraphics[width=1.4\columnwidth]{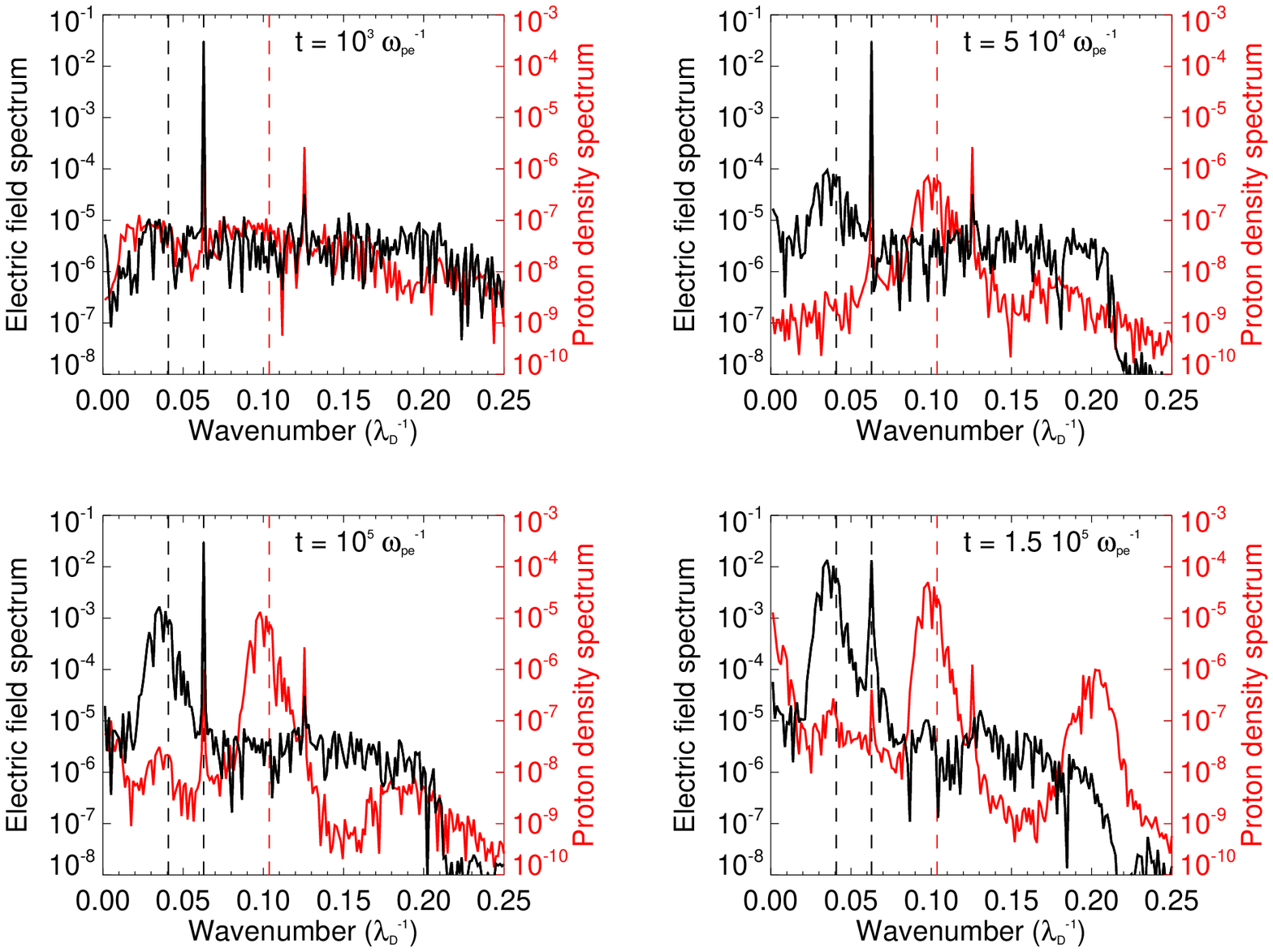}
\caption{Spectrum of both electric field (in black) and ion density (in red) at 4 different times (indicated on each panel), from simulation of LED of a monochromatic Langmuir wave with wavenumber $k_{_{L}} = 0.063$ and initial electric amplitude $E_{_{L}} = 3 \times 10^{-2}$. Dashed lines indicate the expected wavenumbers of the LED waves. Top left panel displays the initial conditions and the bottom right panel shows the spectrum at saturation. Note the presence of the second harmonic of the IAW in this last case.}
\label{fig:spectri}
\end{center}
\end{figure*}

\begin{figure*}
\begin{center}
\noindent \includegraphics[width=1.2\columnwidth]{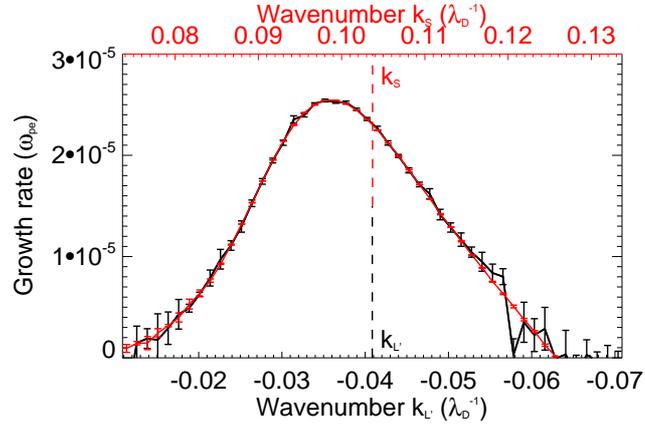} 
\caption{Spectral growth rate for the Langmuir wave $L'$ electric field (in red, $k_{_{L'}}$-axis at the bottom, leftward) and the IAW density fluctuations (in black, $k_{_{S}}$-axis at the top, rightward) generated by the LED of a monochromatic Langmuir wave ($E_{_{L}} = 3 \times 10^{-2}$ and $\Theta = 1$). Both axis are scaled in order to show the resonant condition in wavenumbers: $k_{_{S}} + k_{_{L'}} = k_{_{L}}$. Vertical error bars are 1-$\sigma$ error bars for the fitting of the growth rate. The vertical dotted line indicates the expected wavenumbers for monochromatic LED products.}
\label{fig:spectralgrowthrate}
\end{center}
\end{figure*}

A broad spectrum of daughter waves generated by the instability is clearly seen on Fig.~\ref{fig:spectri}. It shows the electric field and ion density spectrum, black and red lines respectively, at given times. Dashed vertical lines indicate the expected wavenumbers of the LED products as reported in table~\ref{table:expectedwaves}.
In top left panel, $t = 10^3\ \omega_{pe}^{-1}$, the spectrum corresponds to the "initial condition" where only the Langmuir wave at $k_{_{L}} = 0.063\ \lambda_{D}^{-1}$ and its harmonic at $0.13\ \lambda_{D}^{-1}$ are present. 
Then, the growth of the LED produces: (i) a daughter Langmuir waves $L'$ at $k_{_{L'}} \simeq 0.04\ \lambda_{D}^{-1}$ in the electric field spectrum; (ii) an IAW at $k_{_{S}} \simeq 0.10\ \lambda_{D}^{-1}$ in the ion density spectrum. Both are shown in the next two panels, $t = 5 \times 10^4\ \omega_{pe}^{-1}$ (top right panel) and $t = 10^5\ \omega_{pe}^{-1}$ (bottom left panel).
Finally, at $t > 10^{5}$ (bottom right panel) harmonics of the IAW are produced, low-$k$ fluctuations are generated and the LED saturates. 
We need to stress here that during the decay phase, even if the Langmuir wave $L$ is monochromatic, the product waves $L'$ and $S$ are both wave packets with wavenumbers centered on the expected wavenumber. 
This results from the fact that different $k$-channels are available for energy transfer from the Langmuir wave toward its decay products. Indeed Fig.~\ref{fig:spectralgrowthrate} shows the linear growth rate vs. $k$ for different wavenumbers of the daughter Langmuir wave electric field (in black) and IAW density fluctuations (in red) spectrum. Growth rates for both daughter waves are overplotted with $k_{_{S}}$-axis and $k_{_{L'}}$-axis such that $k_{_{S}} + k_{_{L'}} = k_{_{L}}$. As expected, the growth rates are the same for each couple of product waves that verify the wavenumber resonant condition.
The spread in wavenumber is about $\Delta k_{_{L'}} = \Delta k_{_{S}} \simeq 0.025$ for both waves, thereby a relative spectral spread of $\Delta k_{_{S}} / k_{_{S}} \simeq 0.2$ for the IAW and $\Delta k_{_{L'}} / k_{_{L'}} \simeq 0.6$ for the Langmuir product wave. Note that the generation of a large spectrum of product waves limits the spatial coherence of the interaction. This point is discussed in section~\ref{subsection:LEDGrowthRate}. \\

We then compute an 'integrated' (over space) ion density fluctuations $<~\delta~n>$ generated by LED defined by:
\begin{equation} \label{eq:deltan}
 <\delta n>(t) =  \sqrt{ \frac{1}{L_{x}} \int_{0}^{L_{x}} \! (n(t,x) - n_{0})^2 \, dx} 
\end{equation}
with the mean ion density $n_{0} = 1$ in dimensionless units.
We define the saturation level $\delta n_{sat}$ as the maximum value reached by the mean density fluctuations $<\delta n>(t)$ during the simulation.
The growth rate for average density fluctuation is here $\gamma_{_{LED}} = 2.3 \times 10^{-5} \omega_{pe}$, or $\gamma_{_{LED}} =  4.1 \times 10^{-2} f_{_{S}}$ when expressed in term of the IAW frequency, defined by $f_{_{S}} = c_{s} / \lambda_{_{S}} = 5.5 \times 10^{-4}$. This means that the characteristic time scale for the density to grow is about 24 IAW periods, in this case. The saturation level $\delta n_{sat} = 8.7 \times 10^{-5}$, obtained at time $t=1.4 \times 10^5\ \omega_{pe}^{-1}$, is of the order of the expected saturation level defined by the ratio of electric energy to the thermal energy $\delta n_{sat}^{0} = \frac{1}{2} E_{_{L}}^2 / (T_{e}+T_{i})= 2.3 \times 10^{-4}$. \\

The simulation has been repeated for several initial Langmuir wave amplitude, with values in the range $10^{-3}<E_{_{L}}<1$, and for two temperature ratios $\Theta = 0.1$ and $\Theta = 1$. 

We found that the evolution becomes strongly non-linear when the Langmuir wave electric field is "not small", i.e. when $E_{_{L}} \gtrsim 0.3$ for $\Theta = 1$ and $E_{_{L}}\gtrsim 0.2$ for $\Theta = 0.1$. When strong non linear processes take place the evolution is not simply driven by the LED mechanism.
Conversely, for small values of $E_{_{L}}$, the system is mainly driven by the LED process so that we can compute the growth rate for average density fluctuations and the density fluctuations saturation level as previously defined.

\begin{figure*}
\begin{center}
\noindent \includegraphics[width=1.3\columnwidth]{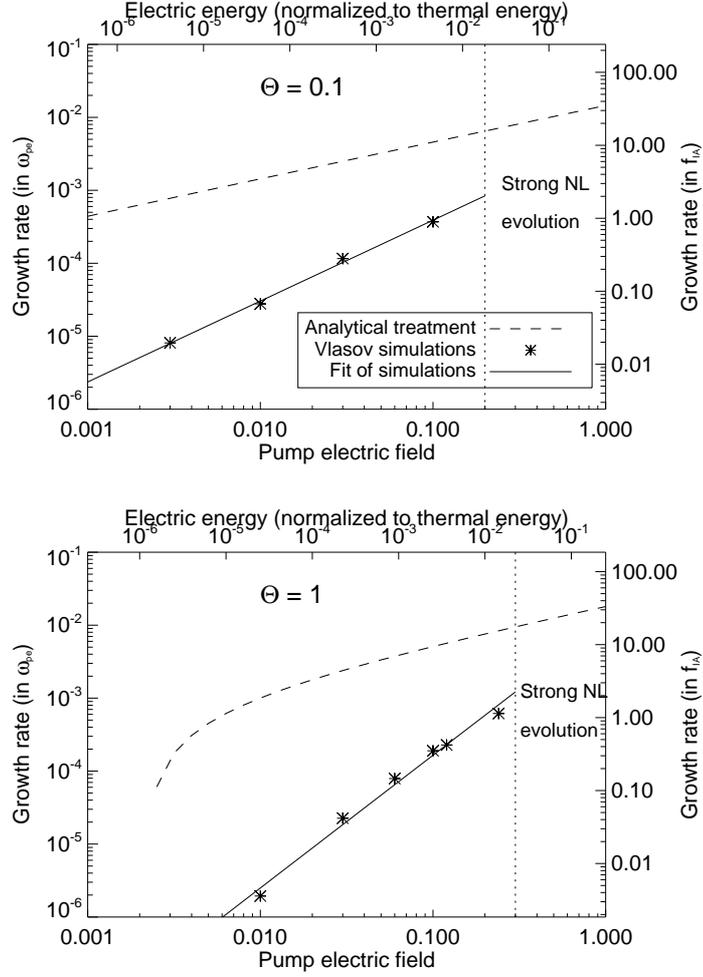} 
\caption{Growth rate for IAW density fluctuations vs. Langmuir wave initial electric field amplitude for two temperature ratios. Top panel: $\Theta = 0.1$; Bottom panel: $\Theta = 1$. The growth rate is expressed in plasma frequency unit $\omega_{pe}$ (left axis) and in IAW frequency $f_{_{IA}}$ (right axis). The full line is a fit of Vlasov simulations (stars). The dashed line shows the analytical growth rate  $\gamma^{analytical}_{_{LED}}$ from \cite{Sagdeev&Galeev1969} when LED-generated IAW remain monochromatic including the linear Landau damping of IAW. The vertical dotted line sets the limits above which strong nonlinear effects occurs before LED starts. }
\label{fig:BilanGrowthRate}
\end{center}
\end{figure*}

The growth rate deduced from simulations $\gamma_{_{LED}}$ of the average density fluctuations $<\delta n>$ deduced from the simulation is displayed in Fig.~\ref{fig:BilanGrowthRate} for the two temperature ratios. As shown before, for too high values of the Langmuir amplitude -- about $E_{_{L}} \simeq 0.3$, corresponding to an electric to thermal energy ratio of $0.1$ -- strong nonlinear effects arise before LED. Also, as expected, the growth rates are lower for $\Theta = 1$ than $\Theta = 0.1$, due to the increase of IAW Landau damping in the first case. Contrary to former results \cite{Umeda&Ito2008PhPl}, we show here that the LED is observed for initial Langmuir waves with electric energy five orders of magnitude lower than the plasma thermal energy. 

An analytical effective growth rate $\gamma^{analytical}_{_{LED}}$ has been overplotted (dashed lines). It is defined as the difference between the analytical full monochromatic case of undamped monochromatic waves in an homogeneous background $\gamma^{th}_{_{LED}}$ \cite{Sagdeev&Galeev1969} and the IAW Landau damping~$\gamma^{Landau}_{_{S}}$:
\begin{equation}
	\gamma^{analytical}_{_{LED}} = \gamma^{th}_{_{LED}} - \gamma^{Landau}_{_{S}}
\end{equation}
where 
\begin{equation}
	\gamma^{th}_{_{LED}} \simeq k_{_{IA}} C_{s} (\frac{\epsilon_{0} E^{2}}{n k_{B} T} \frac{m_{p}}{m_{e}})^{1/4}
\end{equation}
$\gamma^{th}_{_{LED}}$ is of the order of the IAW frequency with solar wind parameters.
The growth rate computed from Vlasov simulations $\gamma_{_{LED}}$ is one to two orders of magnitude lower than $\gamma^{analytical}_{_{LED}}$. The discussion concerning the discrepancy between the analytical monochromatic case and Vlasov simulations is postponed to section~\ref{subsection:LEDGrowthRate}. 

We also looked for the dependence of the growth rate with respect to the mother Langmuir wavevector $k_{_{L}}$. Finally, the growth rate for density fluctuations generated by LED has been fitted by a power law:
\begin{equation} \label{eq:FitGrowthRate}
	\gamma_{_{LED}} = \Gamma \ E_{_{L}}^\alpha k_{_{L}}^\beta
\end{equation} 
The fitting parameters $\Gamma$, $\alpha$ and $\beta$ are given in table~\ref{table:FitValues}. \\

%%%%%%%%%%
\begin{table}
\begin{center}
\begin{tabular}{|c|c|c|c|}
\hline
                                  &   $\Gamma$  &  $\alpha$  &  $\beta$   \\ \hline
	$T_{p} / T_{e} = 0.1$  &      $0.026$    &   $1.11$    &  $0.59$    \\ \hline
	$T_{p} / T_{e} =    1$  &      $0.025$    &   $1.82$    &  $0.30$    \\ \hline
\end{tabular}
\end{center}
\caption{Numerical values for the fit for the growth rate of LED driven density fluctuations in expression~\ref{eq:FitGrowthRate}.} \label{table:FitValues}
\end{table}
%%%%%%%%%%

\begin{figure*}
\begin{center}
\noindent \includegraphics[width=1.3\columnwidth]{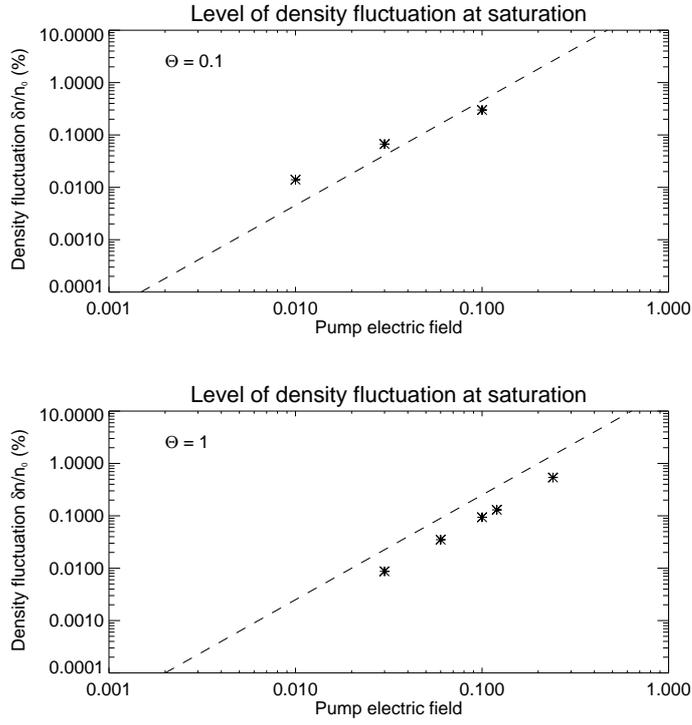} 
\caption{Average density fluctuation at saturation $\delta n_{sat}$ (expressed in $\%$ of mean density) for different Langmuir wave initial electric field amplitude, for both $\Theta = 0.1$ (top panel) and $\Theta = 1$ (bottom panel). The dashed lines show the expected level of saturation $\delta n^{sat}_{0} = (0.5 \times E_{_{L}}^2) / (T_{e}+T_{i})$ defined as the Langmuir electric energy to the total kinetic energy ratio.}
\label{fig:BilanDensityFluctuation}
\end{center}
\end{figure*}

The saturation level of density fluctuation $\delta n_{sat}$ is summarized in figure~\ref{fig:BilanDensityFluctuation} for $\Theta = 0.1$ (top panel) and $\Theta = 1$ (top panel). The expected saturation level $<\delta n^{sat}_{0}>$, expressed as the initial Langmuir electric energy to the total kinetic energy ratio,
\begin{equation}
<\delta n^{sat}_{0}>  = \frac{1}{2} E_{_{L}}^2 / (T_{e}+T_{i})
\end{equation}
is overplotted (dashed lines). The obtained saturation level of density fluctuation is in good agreement with the expected values in both cases. \\

% Resume des resultats
To summarize, the simulations of the Langmuir Electrostatic Decay from an initial monochromatic Langmuir wave have shown that: 
\begin{enumerate}
	\item the threshold for the instability to grow, expressed in term of the Langmuir wave electric energy, is at least 5 orders of magnitude lower than the plasma thermal energy when $0.1 < \Theta < 1$;
	\item the product waves are resonantly generated over a broad range of wavenumbers, naturally producing narrow wave packets;
	\item growth rates of IAW density fluctuations are one to two orders of magnitude lower than the analytical values deduced from the pure monochromatic case;
	\item saturation levels for IAW relative density fluctuations are of the order of the ratio of Langmuir electric energy to the total kinetic energy.
\end{enumerate}

%-------------------------------------------
\subsection{Electrostatic decay of a Langmuir wave packet} \label{section:wavepacket}
In the following, we consider the evolution of a finite-amplitude Langmuir wave packet, generated by the external electric field pump defined in Eq.~\ref{eq:Eext_wavepacket}. 
We recall that, in order to mimic the presence of a low level of small scale ion acoustic turbulence, we introduce a source term of ion acoustic noise. This forcing generates incoherent proton density fluctuations at a level $\delta n / n \sim 10^{-5}$ much smaller than the expected level of density fluctuations generated by LED.

\begin{figure*}
\begin{center}
\noindent \includegraphics[width=1.8\columnwidth]{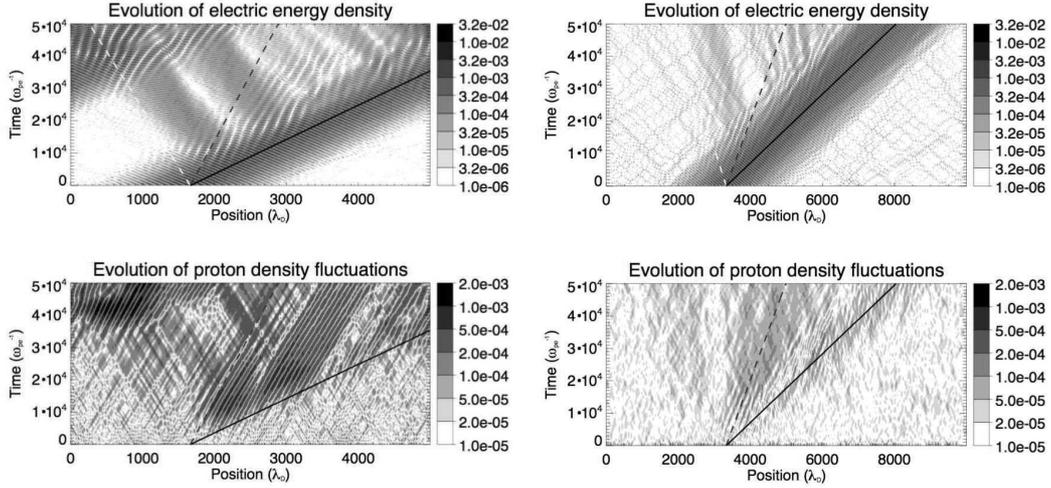} 
\caption{Top panels:  space-time evolution of the electric energy density; Bottom panels: space-time evolution of the ion density fluctuations. Left panels: $\Theta = 0.1$, right panels: $\Theta = 1$. Full lines show the expected group velocities of the Langmuir wave packets while dashed lines displays the ion sound speed.} \label{fig:LEDwavepacket} 
\end{center}
\end{figure*}

Figure~\ref{fig:LEDwavepacket} shows the LED of a Langmuir wave packet with wavelengths centered on $\lambda_{_{L}} = 200\ \lambda_{D}$, a packet width $\Delta = 2000\ \lambda_{D}$, a maximum initial electric field $E_{_{L}} = 6 \times 10^{-2}$, and the two temperature ratio $\Theta = 0.1$ (left panels) and $\Theta = 1$ (right panels). The mother and daughter Langmuir wave packets can be followed in the top panels that show the space-time evolution of the electric field density energy $E(x,t)^2/2$. The Langmuir mother wave packet propagates towards the right and emits backscattered Langmuir wave packets. The bottom panels shows the temporal evolution of ion density fluctuations during the decay of the Langmuir wave. IAW density fluctuations are generated locally (by ponderomotive force from the two Langmuir wave packets beats) and propagate forward at the ion sound speed (dashed line).
When $\Theta = 0.1$ IAW propagate and escape the area where LED develops. 
Conversely, when $\Theta = 1$, IAW density fluctuations are heavily damped as soon as the waves escape the area where LED occurs. Thereby, the waves can be observed only where the mother Langmuir wave decays. Finally, as for the case of monochromatic Langmuir waves, the LED growth rate is lower for a larger value of $\Theta$. \\

\begin{figure*}
\begin{center}
 \noindent \includegraphics[width=1.8\columnwidth]{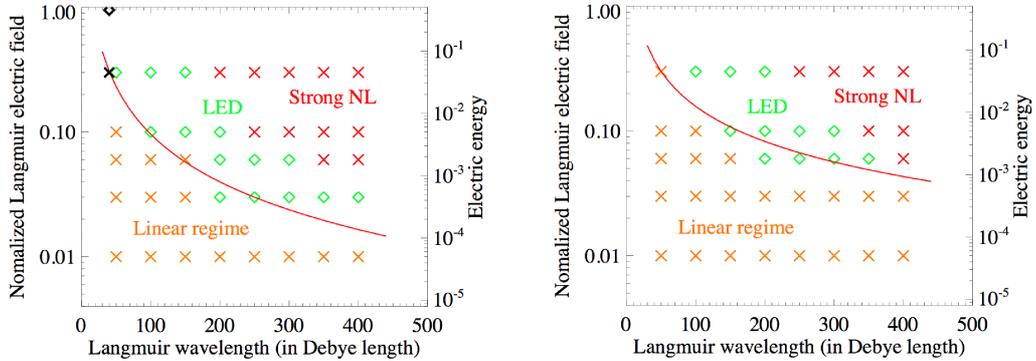} 
\caption{Evolution of a Langmuir wave packet, with different wavelengths centered on $\lambda_{_{L}}$ (bottom axis) but a same packet width of $\Delta=10\ \lambda_{_{L}}$, and different initial electric fields $E_{_{L}}$ (left axis). Left Panel: $\Theta = 0.1$; Right panel: $\Theta = 1$. In green: simulations where LED is observed; In orange: simulations where the available interaction time between the waves is lower than the LED timescale; In red: simulations where strong NL effect arise before/instead of LED. Previous simulation results form \cite{UmedaNPG2007,Umeda&Ito2008PhPl} are overplotted in the left panel with black cross and diamond respectively. The red line is the semi analytical threshold expressed in Eq~\ref{eq:semianalyticalthreshold}.}
\label{fig:wavepacketsummary}
\end{center}
\end{figure*}

The simulation has been repeated for different values of $E_{_{L}}$ and mean wavelength $\lambda_{_{L}}$, with a packet width of $\Delta = 10\ \lambda_{_{L}}$. Figure~\ref{fig:wavepacketsummary} summarizes the results for the evolution of the Langmuir wave packet, each point representing a single simulation. 

During the decay process, the mother wave packet generates a daughter wave packet travelling at a different group velocity. The region of parameters leading to electrostatic decay is displayed with green squares in figure~\ref{fig:wavepacketsummary}. Eventually the two wave packets separate thus stopping the LED process. Therefore, LED is efficient only if the interaction time between the two wave packets is longer than the growth time for the daughter waves. The growth time of the instability is controlled by the Langmuir electric field amplitude: the larger the amplitude, the smaller the growth time (Fig.~\ref{fig:BilanGrowthRate}). The interaction time is controlled by the Langmuir wavelength: the larger the wavelength, the smaller its group velocity and so the larger the interaction time. Thus, for low amplitude and/or short wavelength, the Langmuir wave packet propagates at its group velocity without non-linear interactions. The region of parameters leading to a linear behavior without electrostatic 
decay is displayed with orange crosses in figure~\ref{fig:wavepacketsummary}. 
The efficiency of LED (green) region increases toward the linear (orange) part (i.e. LED is observed for lower $\lambda_{_{L}}$ and $E_{_{L}}$) if temperature ratio $\Theta$ decreases or/and the size $\Delta$ of the mother wave packet increases.
Finally, strong non linear effects dominate the evolution of high amplitude and/or large wavelength Langmuir wave packets, and LED is no more the dominant process. The region of parameters where other non linear effects are dominant is shown with red crosses in figure~\ref{fig:wavepacketsummary}. These strong non linear effects seem to be the signature of strong turbulence: Langmuir collapse and formation of cavitons. This evolution is out of scope of the present study and will be studied in a future work.

The effective threshold for the electrostatic decay of a Langmuir wave packet can be estimated by imposing that the interaction time $\tau_{int}$ is equal to the inverse growth rate of daughter wavepackets $\gamma_{_{LED}}$. The growth rate for LED $\gamma_{_{LED}}$ has been obtained from simulations of monochromatic Langmuir waves LED and fitted in Eq.~\ref{eq:FitGrowthRate}. The interaction time $\tau_{int}$ is evaluated from:
\begin{equation} \label{eq:tau_int}
	\tau_{int} \simeq \Delta / (v^{g}_{_{L}} - v^{g}_{_{L'}})
\end{equation}
where $(v^{g}_{_{L}} - v^{g}_{_{L'}})$ is the difference of group velocity between the two waves that first separate --~here the mother and daughter Langmuir waves~-- and $\Delta$ the packet width of the mother Langmuir wave, which is assumed to be about the length of interaction. 
Using Eq.~\ref{eq:resonantconditions} and $k_{_{S}} \simeq 2\ k_{_{L}}$, Eq.~\ref{eq:tau_int} can be written in normalized units as: 
\begin{equation}
	\tau_{int} \simeq \Delta / (6\ k_{_{L}}).
\end{equation}
Finally, $\tau_{int} = \gamma_{_{LED}}^{-1}$ gives the effective LED threshold of a Langmuir wavepacket with wavevector $k_{_{L}}$ and a packet width $\Delta$:
\begin{equation} \label{eq:semianalyticalthreshold} 
	E^{threshold}_{_{LED}} = \Big( \frac{6\ k_{_{L}}^{1-\beta}}{\Delta \Gamma} \Big)^{1/\alpha}
\end{equation} 
expressed in normalized units. 
The semi analytical threshold is overplotted in Fig.~\ref{fig:wavepacketsummary} (red line) in order to validate its dependency with the mother Langmuir wave vector $k_{_{L}}$. 

\begin{figure*}
\begin{center}
 \noindent \includegraphics[width=1.8\columnwidth]{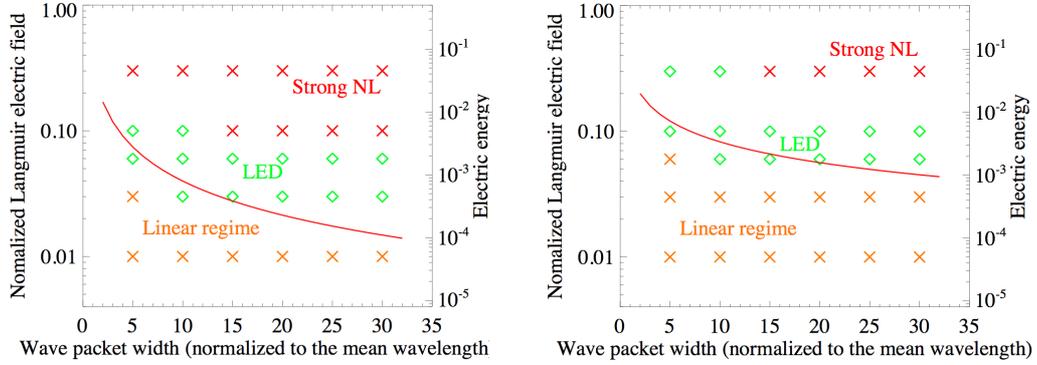} 
\caption{Evolution of a Langmuir wave packet, with a mean wavelength $\lambda_{_{L}} = 200$ but different packet widths $\Delta$ (bottom axis, normalized to $lambda_{_{L}}$), and different initial electric fields $E_{_{L}}$ (left axis). Left Panel: $\Theta = 0.1$; Right panel: $\Theta = 1$. In green: simulations where LED is observed; In orange: simulations where the available interaction time between the waves is lower than the LED timescale; In red: simulations where strong NL effect arise before/instead of LED. The red line is the semi analytical threshold expressed in Eq~\ref{eq:semianalyticalthreshold}.}
\label{fig:wavepacketDelta}
\end{center}
\end{figure*}

We also performed a series of simulations in order to validate the dependency of $E^{threshold}_{_{LED}}$ with $\Delta$. This time the simulation starts with a mother Langmuir wave packet of mean wavelength $\lambda_{L} = 200$ but with different packet width values in the range $5\ \lambda_{L} < \Delta < 30\ \lambda_{L}$, covering the typical range for Langmuir wavepackets in the solar wind, and initial amplitude $E_{L}$. The results are presented in Fig.~\ref{fig:wavepacketDelta}, together with the semi analytical threshold (red line). 
In both cases, Eq.~\ref{eq:semianalyticalthreshold} is in agreement with simulations of LED of localized Langmuir wave packets. 

The threshold decreases when: (i) the ion to electron temperature ratio decreases, since the Landau damping of IAW decreases and their effective growth rate increase; (ii) the mean Langmuir wavelength increases, since Langmuir wave packets with longer wavelengths propagate at smaller group velocities thus increasing the available interaction time with the ion background;
(iii) the width of the Langmuir wave packets increases, since the interaction time between the mother and daughter waves increases.  
The effective threshold obtained by Vlasov simulations and described in this section is compared to observations of LED in the solar wind in the next section.

Finally, in two recent works of 1D-1V Vlasov-simulations, \cite{UmedaNPG2007} and \cite{Umeda&Ito2008PhPl} reported the evolution of beam-excited Langmuir waves. 
Their simulation parameters are $\Theta = 0.1$, a beam speed $v_{beam} = 8\times v_{th,e}$ with a beam temperature equals to the electron core temperature and a beam density ratio of $0.1\%$ and $0.5\%$ respectively. These beams generate Langmuir waves at a phase velocity of about $v_{beam} - v_{th,e} = 7\times v_{th,e}$, i.e. a Langmuir wavelength $\lambda_{L} \simeq 40\times \lambda_{D}$. The amplitude of the Langmuir waves reaches $E_{L} \sim 0.3$ in the first case, $E_{L} \sim 1$ in the second case. They observed the LED process in the second case only.
We have overplotted their results in Fig.~\ref{fig:wavepacketsummary} (black cross and diamond respectively). These previous simulations agrees with Eq.~\ref{eq:semianalyticalthreshold}: for these parameters, the equation predicts that LED of localized Langmuir wave packets should occur for high Langmuir electric energy levels (of the order of unity). \\

To summarize the simulations of the Langmuir electrostatic decay from an initial Langmuir wave packet: 
\begin{enumerate}
	\item we have illustrated how the localization of Langmuir wave packets is crucial for the evolution of Langmuir decay, by limiting the interaction time between mother and daughter waves;
	\item we have shown that the ion acoustic waves are generated locally, where the wave packets interact; then IAW are damped as soon as they escape the region where the mother and daughter Langmuir waves interact when electron and ion temperature are equal, or escape this region when the electron temperature is higher than the ion temperature;
	\item we used results from the Langmuir electrostatic decay of a monochromatic Langmuir wave (section~\ref{section:monochromatic}) to compute a semi analytical threshold for the electrostatic decay of Langmuir wave packets (Eq.~\ref{eq:semianalyticalthreshold}); this semi analytical threshold has been shown to be in agreement with simulations of the electrostatic decay of Langmuir wave packets.
\end{enumerate}

%%%%%%%%%%%%%%%%%%
\section{Discussions} \label{section:discussion}

In the following, we discuss, first, the initialization of LED in the case $\Theta=1$, second, the discrepancy between the LED growth rate obtained from simulations and analytical estimation. 
LED threshold obtained from simulations is then compared to observations. Finally, we show that the saturation of the instability gives an upper limit to the expected level of observed density fluctuations.

\subsection{Initialization of Langmuir electrostatic decay} \label{subsection:LEDinit}

In the case of equal electron and ion temperature, associated with a strong Landau damping of ion acoustic fluctuations, one may wonder whether processes complementary to the resonant interaction of the waves could facilitate the LED. In particular, a decrease (even local) of the IAW Landau damping during the beginning of the LED process could ease the initialization of the instability. Such decrease can happen either through two main processes. (i) Wave-particle interactions due to the trapping of ions in the IAW potential well could modify the ion distribution function by forming a plateau at the ion sound speed which, in turn, could then decrease the Landau damping rate. However, no such plateauing is observed. (ii) The beats of the mother and daughter Langmuir waves, could heat the electrons so that the ion-to-electron temperature ratio decrease locally, leading to a partial suppression of the IAW Landau damping. However the temperature ratio during the simulation has too small variations $\Delta \Theta /
 \Theta < 1 \%$ to really modify the IAW Landau damping. This hypothesis is thus also ruled out. 

Since no such complementary processes are present in the simulations, the initiation of the LED is likely to be caused by the resonant interaction of the waves, that dominates locally the Landau damping of IAW.

\subsection{Growth rate for Langmuir electrostatic decay} \label{subsection:LEDGrowthRate}

We have seen in sect.~\ref{section:monochromatic} that a large discrepancy exists between the growth rate of mean density fluctuations obtained from the simulations, $\gamma_{_{LED}}$, and the analytical one deduced from three monochromatic waves, $\gamma^{th}_{_{LED}}$ \cite{Sagdeev&Galeev1969}\footnote{The influence of the resolution in velocity has been checked, and the results are unchanged as far as the resolution in velocity is not too low. The results are also independent of the size of the box, as long as the box remains larger than the coherence length of the daughter wave packets.}.
The existence of a strong IAW Landau damping $\gamma_{_{Landau}}$, such that $\gamma_{_{LED}} = \gamma^{th}_{_{LED}} - \gamma_{_{Landau}}$ could explain this difference.
Since Landau damping for IAW cannot be found analytically when electron and temperature are close, we performed complementary simulations and measured an IAW Landau damping of $\gamma_{_{Landau}} = -1.9 \times 10^{-5}\ \omega_{pe}$ for $\Theta = 0.1$ (resp. $\gamma_{_{Landau}} = -8.9 \times 10^{-4}\ \omega_{pe}$ for $\Theta=1$), much lower than $\gamma_{th}$. Thus, the hypothesis that $\gamma_{_{LED}} = \gamma^{th}_{_{LED}} - \gamma_{_{Landau}}$ is to be ruled out.

Note that the growth rate measured from simulations already takes into account the effect of the Landau damping. Thus, the difference between $\gamma_{_{LED}}$ and $\gamma^{analytical}_{_{LED}}$ should be carried by the evaluation of $\gamma^{th}_{_{LED}}$.
Actually, the main difference between the analytical treatment of three-wave resonance and Vlasov simulations is that the daughter waves are treated as monochromatic in the first case whereas they are observed to have a non negligible spectral width in the second (see Fig~\ref{fig:spectri} and~\ref{fig:spectralgrowthrate}). 
Simulations have shown that the spectral width of the product waves increases when: 
(i) $\Theta \simeq 1$, in which case the IAW dispersion relation is numerically observed to spread out around the analytical branch; therefore the resonance can occur with $(\omega,k)$-values slightly different from the theoretical expectation; 
(ii) the energy of the Langmuir mother wave increases, in which case the linear approximation for a $\delta$-shaped resonance is no longer valid.
The limited spatial coherence of daughter wave packets implies that their growth is localized where mother and daughter wave packets interact, thus strongly limiting their growth rate. This explains why the growth rate deduced from simulations is much lower than the analytical one.

%-----------------------------------------
\subsection{Threshold of Langmuir electrostatic decay and type~III observations.} \label{section:observations}
% Description of observations:
The TDS observation mode of the S/WAVES experiment on board the STEREO mission \cite{Bougeret&al2007SSR} gives access to {\it in-situ} electric field waveform in 3D with an equivalent spectral resolution up to 60 kHz. In Paper I, we have shown evidence for nonlinear coupling between Langmuir waves at about 10 kHz and ion acoustic waves at about 0.2 kHz. We recall here the global plasma parameters for these observations. The one-hour-average wind speed from WIND/3-DP \cite{Lin&al1995SSRv} is about $V_{SW}~=~315~\mathrm{\ km\ s^{-1}}$. The electron temperature observed by WIND/3DP is $T_e~\simeq~10^{5}~\ K$, and the electron density in the solar wind, estimated from the plasma frequency, is about $n_e \simeq 10^{6} \ \mathrm{m^{-3}}$. From the electron density and temperature, the Debye length is $\lambda_{D} \simeq 20$ m. 

\begin{figure*}
\begin{center}
 \centering \noindent \includegraphics[width=2\columnwidth]{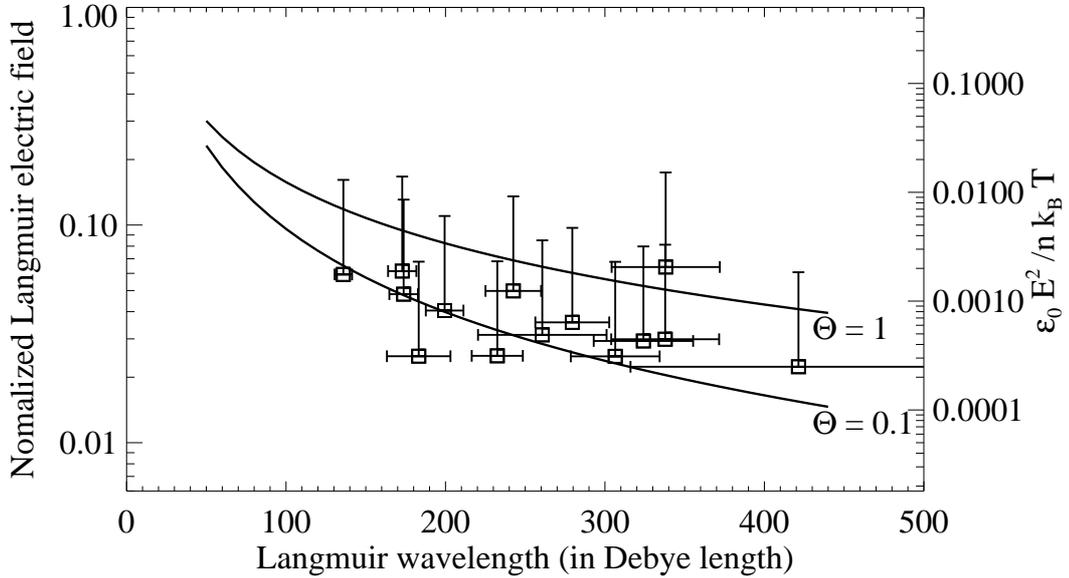}
\caption{Observed Langmuir electric field amplitude -- normalized like the simulations -- vs. wavelength. The threshold for LED computed from Vlasov simulations is overplotted for two values of the temperature ratio $\Theta =0.1$ and $\Theta =1$.} \label{fig:obsVSsimu}
\end{center}
\end{figure*}

Figure~\ref{fig:obsVSsimu} displays the observed value of Langmuir electric field, normalized as described in Eq.~\ref{eq:normalizedelectricfield}, for the whole data set of waveforms where LED has been observed. 
Their wavelength, not directly measured, is evaluated as follow. By taking into account the Doppler-shift caused by the solar wind and noting that the IAW should propagate here anti-sunward, the ion acoustic wavelength reads:
\[ \lambda_{_{S}} = (V_{SW}+c_{s}) / f_{_{S}}^{observed} \]
where the sound speed $c_{s}$ is evaluated by 
\[ c_{s} = \sqrt{k_{B} (T_{e} + T_{p}) / m_{p}},  \] 
where $k_{B}$ is the Boltzman constant, $m_{p}$ the proton mass and $T_{p}$ the proton temperature that remains unknown but is of the order of $T_{e}$ in the solar wind. The mother Langmuir wavelength $\lambda_{_{L}}$ is then about twice the LED-produced ion acoustic wavelength. 

However, caution should be taken when directly comparing this threshold value with the observation. Indeed solar wind type~III Langmuir wave packets are most probably localized in 3D. The observed waveforms are 1D spatial cuts of the real 3D structures and thus only give a lower limit on the width of the wave packet, as well as its maximum amplitude. We assume that the Langmuir wave packets have a 3D gaussian shape, and that the spacecraft crosses it somewhere within a distance from the center of the 3D structure of the order of the full width at half maximum FWHM. Therefore a realistic range for the maximum electric field $E_{L}^{real}$ reached by the 3D wave packet is calculated as: $E_{L}^{obs} \lesssim E_{L}^{real} \lesssim e\ E_{L}^{obs}$, where $E_{L}^{obs}$ is the observed level of Langmuir electric field. Data are plotted with such error bars.

The LED threshold expressed in Eq.~\ref{eq:semianalyticalthreshold} is overplotted for two values of the ion to electron temperature ratio $\Theta = 0.1$ and $\Theta = 1$, using computed values from table~\ref{table:FitValues}.
The observed electric field amplitudes are of the order of the threshold computed from Vlasov-Poisson simulations -- in the case of a localized wave packet and $\Theta\approx 1$ -- confirming that the development of LED is compatible with the observed events.

%-----------------------------------------
 \subsection{Saturation of LED: expected level of observed density fluctuations} \label{subsection:DensityFluctuations}

In our simulations, for initial Langmuir wave packets in the range of amplitudes that correspond to observations of Langmuir waves during type~III events, the IAW-like density fluctuations associated with the LED do not reach the saturation level. Indeed, the mother and Langmuir wave packets that propagate at different group velocities separate before the saturation stage, thus stopping the growth of the IAW. 
This is confirmed in the observations, since most of the observed waveforms show: (i) mother and daughter Langmuir waves with different energies; (ii) IAW with no harmonics. The saturation level $<\delta n^{sat}_{0}> / n_{0} = E_{_{L}}^2 / (T_{e}+T_{i})$ should thus be an upper limit on the expected level of density fluctuations on observations of LED during type~III.

Density fluctuations at the frequencies considered here can be measured in space from the variations of the floating potential of the spacecraft \cite{Pedersen1995AnGeo} when it crosses the region where LED is observed. The voltage signal observed on STEREO/Waves waveforms with monopole antennas channels contains information on the floating potential of the spacecraft. This remains however to be calibrated. Knowing the saturation level of IAW during LED, and given the calibration, we could check whether the observed level of density fluctuations is consistent with our simulations.

%%%%%%%%%%%%%%%%%%
\section{Conclusion}

% Resume des resultats (repondre aux question de l'intro)
%--------------------------------------------------------------------------------
In order to study the origin of electromagnetic radio emissions during type~III bursts, we have reported in this paper 1D-1V Vlasov-Poisson simulations of the Langmuir Electrostatic Decay. The simulations have been done in typical solar wind conditions: ratio of the electron to ion temperature from 0.1 to 1, mother Langmuir wavelengths typical of those observed during type~III events and, most important, by considering localized Langmuir wave packets. 
The main results are the following:
\begin{enumerate}
	\item Langmuir electrostatic decay develops even when the electron and ion temperatures are close. Its threshold, when considering a monochromatic wave, is at least five orders of magnitude lower than the plasma thermal energy when $0.1 < \Theta < 1$.	
	\item Langmuir electrostatic decay resonantly generates daughter waves over a broad range of wavenumbers, naturally leading to narrow wave packets. This limits the length of coherence which is why the growth rate is one to two orders of magnitude lower than the analytical values deduced from a pure monochromatic case.
	\item The behavior of daughter ion acoustic waves depends on the temperature ratio. Ion acoustic waves can escape the region where the resonant coupling takes place and propagate when electron temperature is higher than proton temperature. Conversely, they are damped as soon as they escape the resonant coupling area when temperatures are equal.
	\item We confirm that the saturation level for IAW density fluctuations is of the order of the ratio of Langmuir electric energy to the total kinetic energy. However, for the range of amplitudes that correspond to observations, the IAW-like density fluctuations associated with the LED should not reach the saturation level because (i) the mother and Langmuir wave packets that propagate at different group velocities separate before the saturation stage, (ii) harmonics of the IAW are seen at saturation in the simulations but, to our knowledge, not in Type~III observations. 
	\item Finally, an effective threshold has been obtained (Eq.~\ref{eq:semianalyticalthreshold}) for localized Langmuir wave packets and compared to STEREO/WAVES observations. The observed Langmuir electric field during type~III reported in \cite{Henri&al2009JGR} is in the range of LED effective threshold computed from Vlasov simulations, thus confirming the interpretation of these observed electric field waveforms in term of the LED of type~III beam-driven Langmuir waves. \\
\end{enumerate}

% Ouverture sur problemes pas resolus / nouveaux:
%-------------------------------------------------------------------------
The physical mechanism responsible for the generation of electromagnetic radio waves during type~III burst is still under study. The process described by \cite{ginzburg58} and leading to the generation of type~III radio emission at twice the local plasma frequency is a two step process. (i) First the beam-driven Langmuir wave decays through LED, (ii) then the mother and daughter Langmuir waves coalesce to generated the electromagnetic wave. 

Up to now, observations have shown that resonant coupling between Langmuir waves and ion acoustic waves does occur during a type~III burst \cite{Henri&al2009JGR}. The present paper also shows that Langmuir electrostatic decay may occur in solar wind conditions and that the threshold is reached by the observed Langmuir electric amplitude. These two complementary studies thus confirms step (i) does occur.

Langmuir Electrostatic Decay generates two counter-propagating Langmuir waves in opposite direction, but step (ii) requires obliquely propagating Langmuir waves for waves to couple and produce the transverse electromagnetic wave. Does the density inhomogeneities scatter the Langmuir waves enough to introduce a significant perpendicular component to their wave vector? An important step would be to study the interaction of Langmuir waves with an inhomogenous background, in order to check this hypothesis. 
Finally, the coalescence of counter-propagating Langmuir waves is hard to observe because the signal would be hidden by the generation of the harmonic of the mother Langmuir wave. Simulations of step (ii) with input from the observations could at least show whether the coalescence process could occur in the solar wind.

%--------------------------------------------------- ANNEX -----------------------------------------------------
\appendix

\section{Appendix: Generation of the ion density noise} \label{appendix:Eext_ions}

The continuous injection of ion density noise is driven by an external fields $E_{ext}^{noise}$ added to the ion dynamics (Eq~\ref{eq:vlasovions}). This driver is defined by:
\begin{equation} \label{eq:forcingions}
E_{ext}^{noise}(x,t) = E_{ext}^{ions, max} \frac{ \sum_{\lambda} cos(2 \pi x / \lambda) cos(\omega_{\lambda} t + \Psi_{\lambda}^{''}(t)) } { | \sum_{\lambda} cos(2 \pi x / \lambda) cos(\omega_{\lambda} t + \Psi_{\lambda}^{''}(t)) |} 
\end{equation} 
It introduces a flat spectrum for wavelength over the range $50 < \lambda < 1000$. $E_{ext}^{noise}(x,t)$ is normalized in order to have a maximum amplitude of $E_{ext}^{ions, max} = 1 \times 10^{-5}$. The frequencies $\omega_{\lambda}$ are chosen to satisfy the dispersion relation of IAW, $\omega_{\lambda} = (2 \pi / \lambda) c_{s}$ with $c_{s} = \sqrt{(1 + \Theta)/\mu}$ the ion sound speed. 
The phases $\psi_{k}''(t)$ have a step-like variation, constant over a time interval $\delta t$. At the end of each interval, they are independently drawn from a uniform distribution.
This way, the generation of ion acoustic noise is made of a succession of time coherent forcing for time intervals of duration $\delta t = 500 \times \omega_{pe}^{-1}$ (about 80 plasma oscillations). This means that for an IAW of wavelength $\lambda_{IA} = 100$, the forcing lasts $1/20^{th}$ of a period, much shorter than the typical IAW oscillation time in order to generate an incoherent noise. This forcing thus generates density fluctuations much smaller than the level of density fluctuations generated by LED in our simulations.

\section{Appendix: Generation of the initial Langmuir wave} \label{appendix:Eext}

In this appendix, we describe the driver~$E_{ext}^{init}$, added in the Vlasov equation for electrons (Eq.~\ref{eq:vlasovelectrons}). This 'external' electric field controls the generation of the initial Langmuir wave. It acts as a driver that resonantly grow a Langmuir wave propagating in only one direction, with the desired spectrum and electric field amplitude.
In section~\ref{section:monochromatic}, a monochromatic Langmuir wave of wavelength~$\lambda_{_{L}}$ and amplitude~$E_{_{L}}$ is resonantly excited by the external electric field pump by defining $E_{ext}^{init} = E_{ext}^{(1)}$:
\begin{equation} \label{eq:Eext_monochromatic}
E_{ext}^{(1)} (x,t) = E_{0}^{(1)} (t) \ cos(k_{0} x - \omega_{0} t)
\end{equation}
In section~\ref{section:wavepacket}, a Langmuir wave packet of mean wavelength~$\lambda_{_{L}}$, packet width~$\Delta$ and maximum amplitude~$E_{_{L}}$ is resonantly excited by defining $E_{ext}^{init} = E_{ext}^{(2)}$:
\begin{equation} \label{eq:Eext_wavepacket}
E_{ext}^{(2)} (x,t) = E_{0}(t) \ cos(k_{0} x - \omega_{0} t)  \ exp\bigg(-\Big(\frac{x-x_{0}-v_{_{L}}^{g} t}{0.5 \Delta}\Big)^2\bigg)
\end{equation}
In both cases, the pump wavevector $k_{0} = 2 \pi / \lambda_{_{L}}$ and the frequency $\omega_{0} = \sqrt{1 + 3 k_{0}^2}$ are chosen to satisfy the Langmuir dispersion relation, in order to generate a Langmuir wave at the desired wavelength $\lambda_{_{L}}$. 
The width of the wave packet is set to $\Delta = 10\ \lambda_{_{L}}$. The pump wave packet moves at the Langmuir group velocity $v_{_{L}}^{g} = 3 k_{0} / \omega_{0}$ from its initial position $x_{0} =  L_{x}/3$. 
The time-dependant external electric field amplitude $E_{0}(t)$ scales on the desired Langmuir ampitude $E_{_{L}}$: 
\begin{equation} \label{eq:EextAmp}
E_{0}^{(1)} (t) = \eta E_{_{L}} (1 - exp(- t / \tau)). 
\end{equation}
In the monochromatic case (resp. wave packet case), the parameter $\tau$ is set to $30\ \omega_{pe}^{-1}$ (resp. $100\ \omega_{pe}^{-1}$) so that the external electric field amplitude smoothly increases during a few Langmuir oscillations, in order to avoid the generation of other plasma waves induced by step-like switch of the pump. The parameter $\eta$ is set experimentally to $5 \times 10^{-2}$ (resp. $10^{-2}$), so that the characteristic time scale to resonantly grow the Langmuir wave (resp. wave packet) is large compared to its oscillation time scale, but small compared to the decay time scale. So, the resonant generation of the Langmuir wave (resp. wave packet) does not interfere with the LED mechanism. \\

%--------------------------------------------------- End of text --------------------------------------------------
\subsection*{Acknowledgments} We are grateful to the italian super-computing center CINECA (Bologna) where part of the calculations where performed. We also acknowledge Dr. C. Cavazzoni for discussion on code performance. \\

\bibliographystyle{unsrt}
% \bibliography{/home/phenri/Documents/COMPTE_RENDUS/mabiblio}

%\bibliographystyle{agu}
%\bibliography{../../../mabiblio}

%%%%%%%%%%%%%%%%%%% Bibliography %%%%%%%%%%%%%%%%%%
% Mettre ici fichier bibliographie (dernier moment)

%######### Graphiques ##################

%%%%%%%%%%
%%%%%%%%%%

%%%%%%%%%%
%%%%%%%%%%

%%%%%%%%%%
%%%%%%%%%%

%%%%%%%%%%
%%%%%%%%%%

%%%%%%%%%%
%%%%%%%%%%

%%%%%%%%%%
%%%%%%%%%%

%%%%%%%%%%
%%%%%%%%%%

%%%%%%%%%%
%%%%%%%%%%

%----------------------------------------------------------------------------------------------------------------------
\end{document}